\documentclass[12 pt, fleqn]{scrartcl}
\usepackage{amsfonts,amssymb,amsmath}
\usepackage{graphicx}
\renewcommand{\Im}{\mathop{\mathrm{Im}}\nolimits}

\newcommand{\grad}{\mathop{\mathrm{grad}}\nolimits}

\begin{document}
\begin{flushright}
ITEP-TH 39/11\\
\end{flushright}

\vskip 10mm
\begin{center}
{\Huge{\bf Monopole solutions to the Bogomolny equation as three-dimensional generalizations of the Kronecker series}
}\\
\vspace{15mm}

K. Bulycheva \\

\vspace{5mm}
{\sf Institute for Theoretical and Experimental Physics, Moscow, Russia} \\

{\sf Moscow Institute for Physics and Technology, Dolgoprudny, Russia} \\

{\em e-mail: bulycheva@itep.ru}\\

\vspace{15mm}
{\bf \abstractname}
\begin{abstract}
The Dirac monopole on a three-dimensional torus is considered as a solution to the Bogomolny equation with non-trivial boundary conditions. The analytical continuation of the obtained solution is shown to be a three-dimensional generalization of the Kronecker series. It satisfies the corresponding functional equation and is invariant under modular transformations.
\end{abstract}

\end{center}



\section{Introduction}

\paragraph{The Bogomolny equation.}
In the most general formulation the monopole solution in the Yang-Mills theory can be defined as the generalization of the Dirac monopole \cite{dirac, dirac2} in electrodynamics to the non-abelian gauge group. The monopole is a classical solution in four-dimensional theory invariant under translations along one of the coordinates \cite{kronheimer,atiyah,mon,rubakov,shnir}. In his famous paper \cite{bogomolny} E.\,Bogomolny suggested a description of monopoles in terms of the solutions to system of first-order equations in the three-dimensional space. In general this space can be replaced by a three-dimensional manifold $\mathcal{M}$ (possibly with boundary) equipped with metric. There is a  $G$-bundle over $\mathcal{M}$, a vector bundle  $E_V$ associated with the $G$-bundle, and an adjoint bundle $\hbox{End}E_V$ with the connection $A\in\Omega^1(\mathcal{M},\hbox{Lie}(G))$. The Higgs field $\phi$ is the section of the adjoint bundle, $\phi\in\Omega^0(\mathcal{M},\hbox{End}E_V)$. Then the Bogomolny equations connect the Higgs field and the curvature,

 \begin{equation}
  \label{Bogomolny} F=\star D\phi,
 \end{equation}

here $F=dA+A\wedge A$ is the curvature of the connection, $\star$ is the Hodge operator, $D$ is the covariant derivative, or $d+A$ in the adjoint. This system is gauge-invariant,

 \begin{equation}
  \label{gauge1} A\rightarrow f^{-1}Af+f^{-1}df, \ \ \ \phi\rightarrow f^{-1}\phi f .
 \end{equation}

The Bogomolny equation can be derived from the self-duality equation $F=\star F$ by means of the dimensional reduction. For example, in the work of Kronheimer \cite{kronheimer} a monopole is considered as an instanton which is invariant under $U(1)$-transformations. Note that in case  $G=U(1), \mathcal{M}=\mathbb R^3$ the Bogomolny equation reproduce the Dirac equation for monopole. Indeed, applying the $D$ operator to the both sides of (\ref{Bogomolny}) leads to the Laplace equation:
 \begin{equation}
  \label{Laplace} \Delta \phi=c_0\delta^{(3)}(\vec r).
 \end{equation}

The $\delta$-function in the r.h.s. reflects the wish to consider the solutions which have the $\sim\frac{1}{r}$ singularity in the given point (which is chosen to be the origin)\footnote{Also it can be said that the Dirac monopole is the solution to the Laplace equation $\Delta \phi=0$ on the manifold $\mathbb R^3/\{0\}$ which has the asymptotic behaviour $\phi\sim\frac{1}{r}$ in $\vec r=0$ and
$\phi\rightarrow0$ in $r\rightarrow\infty$.}. This solution which decreases at infinity is $\phi=\frac {c_0}{r}$. The magnetic field is described with $D\phi=\grad \phi$.

\paragraph{Hecke operators and monopoles.}

This work was motivated by the results of \cite{witten} and \cite{article}. In the paper \cite{witten} the connection between the geometric Langlands program and the four-dimensional Yang-Mills theory compactified to the complex curve $C$ was discussed. The results of \cite{witten} mean in particular that in case $\mathcal{M}=C\times\mathbb R$ the emergence of the monopole solution to the equations (\ref{Bogomolny}) can be treated as the application of the Hecke operator in the sections of the (holomorphic) bundle ${\tilde
E}_V=E_V\left.\right|_{C\times\{0\}}$ over $C$. Let $z,\bar z$ are local complex coordinates on $C$ and $y$ be the coordinate on $\mathbb R$. The system (\ref{Bogomolny}) in components is as following:

 \begin{equation} \label{sys_t2} \left\{
  \begin{array}{l}
  \partial_z A_{\bar z}-\partial_{\bar z}A_z+[A_z,A_{\bar z}]=\frac {ig}2 \left(\partial_y \phi+[A_y,\phi]\right),\\
  \partial_y A_{z}-\partial_{z}A_y+[A_y,A_{z}]=i \left(\partial_z \phi+[A_z,\phi]\right),\\
  \partial_y A_{\bar z}-\partial_{\bar z}A_y+[A_y,A_{\bar z}]=-i \left(\partial_{\bar z} \phi+[A_{\bar z},\phi]\right).\\
  \end{array}
  \right.
 \end{equation}

Following \cite{article}, the gauge is chosen as \footnote{The condition $A_{\bar z}=0$ fixes not all the gauge degrees of freedom, because the holomorphic gauge transformations leave it invariant. The system (\ref{sys_t2}) under this condition contains the relation $\partial_{\bar z}(A_y-i\phi)=0$. This means that by means of the holomorphic gauge transformations the condition $A_y=i\phi$ can be satisfied.}: $A_{\bar z}=0$, $A_y=i\phi$. Finally the system becomes:

 \begin{equation}
  \label{Bogom_complex} \left\{\begin{array}{lcr}
  \partial_{\bar z}A_z=-\frac{ig}2 \partial_y\phi,\\
  \partial_yA_z=2i\partial_z\phi-2i\left[A_z,\phi\right].
  \end{array}
  \right.
 \end{equation}

The Hecke operator (or the modification of the bundle) acts in $\tilde V$ and changes its degree $\deg\tilde V$, i.e. adds zeroes and poles to the sections of the determinant bundle $\det \tilde V$. Consider for simplicity the abelian case on $\mathcal{M}=\mathbb C\times\mathbb R$. Then the holomorphic gauge transformation (\ref{gauge1}) $f=z^m$, $m\in\mathbb Z$ acts on the connection as $A_z\rightarrow A_z+\frac{m}{z}$. On the other hand, $\phi=\frac{c_0}{r}=\frac{c_0}{\sqrt{y^2+z\bar z}}$, and (\ref{Bogom_complex}) implies that the solution for $A_z$ with boundary condition $A_z|_{\infty}=0$ is:

 \begin{equation}
  \left\{
  \begin{array}{ll}
  A_z=-ic_0\left(\frac 1z \frac y{\sqrt {y^2+z\bar z}}-\frac 1z\right),&y>0,\\
  A_z=-ic_0\left(\frac 1z \frac y{\sqrt {y^2+z\bar z}}+\frac 1z\right),&y<0.\\
  \end{array}
  \right.
 \end{equation}

The connection jumps at $y=0$ by $-2ic_0\frac 1z$. Hence $c_0=\frac {im}2$, $m\in\mathbb Z$, and that corresponds to the 'quantization' of the monopole charge \cite{witten,article}. In this way we have demonstrated the connection between the monopole solution and the Hecke operator. In general the modification of the bundle is parameterized by the elements of the coweight lattice for the $G$ group \cite{witten,article,LOSZ}. According to Kronheimer \cite{kronheimer} the monopole is the instanton in $\mathbb C^2$ which is invariant under the action of $U(1)$ group. The $U(1)$ action has the fixed point $z_1=z_2=0$. If the action of $U(1)$ group is lifted to the $G$-bundle over $\mathbb C^2$ then $U(1)$ acts in particular in the fiber over the fixed point and defines the homomorphism $U(1) \rightarrow G$, or cocharacter of $G$. Hence the action of the $U(1)$ group which makes the reduction from the instanton to the monopole generates the set of coweights defining the modification \cite{witten,article,LOSZ}.

\paragraph{Hitchin systems and monopoles.}

In the N.\,Hitchin's approach to the integrable systems \cite{hitchin} the latter naturally arise on the moduli spaces of the holomorphic bundles. The modification changes the degree (or more precisely the characteristic class \cite{LOSZ}) of the bundle and connects systems of different types, such as for example Calogero systems and Euler-Arnold top \cite{levin}. In the case of the moduli space for flat connections similar results provide connection between different non-autonomous generalizations of the Hitchin systems which are non-linear equations of the Painlev\'e type \cite{LOZ2}. The connection between monopoles and modifications suggests that there exists a non-abelian solution which connects the parameters of the systems at boundary (at $y=\pm\infty$). A certain class of the non-abelian is known. For example the Nahm construction \cite{corrigan} allows to solve the Bogomolny equations for the $SU(N)$ group. One else possible way is to use the methods of the soliton theory \cite{atiyah}. The Bogomolny equation can be represented as a zero curvature condition for some differential operators in the same way as the Yang-Mills-Higgs equation \cite{ward}. However the explicit construction of non-abelian solution with the desired asymptotic behaviour is not known.

\paragraph{The aim of the paper.} 
The current paper suggests a generalization of the result obtained in \cite{article}. In that paper the authors considered the scalar Bogomolny equation on the $\Sigma_\tau \times \mathbb R$ manifold, where $\Sigma_\tau$ is the elliptic curve realized as a quotient of the complex plane $\mathbb C$ modulo the period lattice $\Gamma_\tau$, which generates the fundamental parallelogram.

In the current paper the scalar Bogomolny equation is considered on the  $\mathbb T^2 \times \mathbb R$ and $\mathbb T^3$ manifolds. The aim is to describe the Green function on these manifolds and to study their modular properties. While investigating the solution on the three-dimensional torus it is convenient not to fix the holomorphic gauge. We begin with describing a periodic Green function for the Laplace equation on $\mathbb T^3$. To do so, we realize $\mathbb T^3$ as ${\mathbb R}^3/\Gamma_3$ where $\Gamma_3$ is a three-dimensional lattice generated by ${\vec\gamma}_i$, $i=1,2,3$ vectors. Then the solution to the given problem is equivalent to the solution of (\ref{Laplace}) on $\mathbb R^3$ with the following boundary condition:

 \begin{equation}
  \label{bc} \phi\left(\vec x+\vec\gamma\right)=\phi\left(\vec x\right),\,\,\,\vec\gamma\in\Gamma_3.
 \end{equation}

The $\delta$-function is the sum of characters of the lattice group $\Gamma_3$ (\ref{char}):

 \begin{equation}
  \delta\left(\vec x\right)=\sum_{\vec n \in \mathbb{Z}^3}\chi_3 \left(\vec x, \vec n\right).
 \end{equation}

Averaging gives the following solution: $\phi\left(\vec x\right)=c_0\sum\limits_{\vec\gamma\in\Gamma_3}\frac 1{\left|\vec x+\vec\gamma\right|}$. However this series diverges. To regularize the series, we introduce the parameter of the analytical continuation. Consider the generalized Laplace equation with the pseudo-differential operator $\Delta^{2s}$ and the corresponding Green function:

 \begin{equation}
  \label{Laplace_gen} \Delta^{2s} \tilde\phi_s\left(\vec x\right)=c_0\delta\left(\vec x\right), \text{where} \ \ \tilde
  \phi_s\left(\vec x\right)=c_0\sum_{\gamma\in\Gamma}\frac 1{\left|\vec x+\gamma\right|^s}.
 \end{equation}

We also introduce the metric $M^TM$ in the Laplace operator (in the case (\ref{Laplace}) the metric is  $M_{ij}=\delta_{ij}$) $\Delta=\vec \partial ^T \left(M^{-1}\right)^TM^{-1} \vec \partial$, where $\vec \partial=\left(\partial_1, \partial_2,\partial_3\right)^T$ and consider quasi-periodic boundary conditions:

\begin{equation}
 \label{cond} R_3\left(s, M, \vec x+\vec \gamma_i, \vec \xi\right)=e^{-2\pi i (w\vec \alpha_i \cdot \vec\xi)}R_3\left(s, M,
 \vec x, \vec \xi\right),
\end{equation}

where $\vec\alpha_i$ are the generating vectors of the dual lattice and $w \in SL(3,\mathbb Z)$ parameterizes quasi-periodic boundary conditions. Then the solution to the equation (\ref{Laplace_gen}) with the boundary conditions (\ref{cond}) is the following:

 \begin{equation}
  \label{R_def} R_3\left(s, M, \vec x, \vec \xi\right)=c_0 \sum_{\vec n \in \mathbb Z^3} \frac{\chi_3\left(\vec {\tilde n},
  \vec \xi\right)}{\left|\vec x+\vec \gamma\right|^s}= c_0 \sum_{\vec n \in \mathbb Z^3}\frac{\chi_3\left(\vec {\tilde n}, \vec
  \xi\right)}{\left(\left(\vec x^T+\vec n^TL_3\right)MM^T\left(\vec x+L_3^T\vec n\right)\right)^s},
 \end{equation}

where  $\vec{\tilde n}=w\vec n$ and $L_3$ is the matrix of the lattice (\ref{l2}). Using the definition of the $\Gamma$-function, we represent (\ref{R_def}) in the integral form\footnote{The similar function can be introduced also for two-dimensional torus:
 \begin{equation}
  \label{I_2} I_2\left(s,M,\vec x, \vec \xi\right)=c_0\sum_{\vec n} \int\frac{dt}t t^s e^{-t\left(\vec x^T+\vec
  n^TL_2\right)M^TM\left(\vec x+L_2^T\vec n\right)} \chi_2\left(\vec n, \vec \xi\right).
 \end{equation}
It is not a solution to the two-dimensional Laplace equation, but we need it while considering the $\mathbb{T}^2\times
\mathbb{R}$ case.}:
\begin{equation}
  \label{I_def} I_3\left(s,M,\vec x, \vec \xi\right)=\Gamma(s) R_3\left(s,M,\vec x, \vec \xi\right)= c_0\sum_{\vec
  n}\int\frac{dt}t t^s e^{-t\left|\vec x+\vec \gamma\right|} \chi_3\left(\vec {\tilde n}, \vec \xi\right).
 \end{equation}
 
The main result of the current work is 

{\bf Theorem.} \emph{The expression (\ref{R_def}) gives the Green function of the generalized three-dimensional Laplace equation (\ref{Laplace_gen}) with quasi-periodic boundary conditions (\ref{cond}), which satisfies the following functional equation:}

\begin{equation}
R_3\left(s,M,\vec x, \vec \xi\right)=\frac{\Gamma\left(\frac 32 -s\right)\pi^{2s-\frac 32}}{\Gamma(s)\det L_3M}e^{-2\pi i
\vec x^TL_3^{-1}\left(L_3^T\right)^{-1}\vec \xi}R_3\left(\frac 32-s,\left(L_3^TL_3\right)^{-1}M,\vec\xi,\vec x\right)
\end{equation}

\emph{and is invariant under modular transformations generated by the elements of the double coset $SL(3,\mathbb Z)\backslash SL(3,\mathbb R)/SO(3,\mathbb R)$.}

The proof follows from Lemma 1 and Lemma 2 given below and the results of the section 3 about modular transformations. We find the Green functions on $\mathbb T^3$ and pass to the limit of $\mathbb T^2\times \mathbb R$ considered in the work \cite{article}. Note that the acquired Green function is the generalization of the Kronecker series by the construction \cite{weyl}.

\section{Functional equation}
\label{fn_eq}

The relation (\ref{R_def}) implies that $R_3$ diverges in the case of $s=\frac 12$, however converges when $s$ is large. To provide the analytical continuation of $I_3$ we use the Poisson summation formula,

 \begin{equation}
\label{Poisson}
\sum_{\vec \gamma \in \Gamma}f\left(\vec \gamma +\vec x\right)=
\sum_{\vec g \in \Gamma^\vee}\hat f \left(\vec g\right)e^{2\pi i \left(\vec x, \vec g\right)}.
 \end{equation}

Here $\Gamma^\vee$ is the dual lattice, $\hat f$ is the Fourier transformation for $f$. Due to the integral form of (\ref{I_def}) all the integrals in the Fourier transformation are Gaussian and the shape of the expression after integration is almost the same. This trick allows to obtain the analytical continuation from $s=\frac 12$ to $s=1$.

{\bf Lemma 1.}\cite{weyl} \emph {The function (\ref{I_2}) satisfies the following functional equation:}
\begin{equation}
\label{2d_eq}
I_2\left(s, M, \vec \xi, \vec x\right)=
\frac {\pi^{2s-1}}{\det M \det^{2s}L_2}I_2\left(1-s, M^{-1}, \vec x, \vec \xi\right)e^{2 \pi i \frac {\vec x^T \sigma \vec \xi}{\det L_2}}.
\end{equation}

\underline {Proof.} Consider the Fourier transformation for (\ref{I_2}). After Gaussian integration and rescaling of $t$ we obtain:

\begin{multline}
\label{d2}
\hat I_2\left(s,M,\vec \xi, \vec x\right)=c_0\int d^2n\sum_{\vec \nu \in \mathbb Z^2}\int
\frac{dt}t t^s e^{-t\left({\vec x}^T+{\vec n}^TL_2\right)M^TM\left({\vec x}+L_2^T{\vec n}\right)}
e^{2\pi i {\vec n}^T L_2^{\vee T} {\vec \xi}}e^{-2\pi i {\vec \nu}^T \vec n}=\\
\frac {c_0}{\det L_2M}\sum_{\vec \nu \in \mathbb Z^2}\int \frac{dt} t t^s \frac \pi t
e^{-\frac{\pi^2}t\left({\vec \xi}^TL_2^{\vee}\left(L_2^T\right)^{-1}-{\vec \nu}^T
\left(L^T\right)^{-1}\right)M^TM\left(L_2^{-1}L_2^{\vee T}\vec \xi-L_2^{-1}{\vec \nu}\right)}
e^{2\pi i {\vec \nu}^T\left(L_2^T\right)^{-1}\vec x}\\e^{-2\pi i {\vec x}^T L_2^{-1}L_2^{\vee T}\vec \xi}=
c_0\sum_{\vec \nu \in \mathbb Z^2}\int \frac{dt} t t^s \frac \pi t
e^{-\frac{\pi^2}t\left(\frac 1{ab}{\vec \xi}^T\sigma^{-1}-
\frac1 {ab}\left({\vec \nu}^T\sigma\right)L_2\sigma^{-1}\right)\left(M^{-1}\right)^TM
\left(\frac \sigma {ab}\vec \xi-\frac \sigma {ab}L_2^T\sigma^T{\vec \nu}\right)}\\
e^{2\pi i \left(-{\vec \nu}^T \sigma^{T}\right)L_2^{\vee T}\vec x}
e^{-2\pi i {\vec x}^T L^{-1}L^{\vee T}\vec \xi}\frac 1{\det ML_2}=
\frac {\pi^{2s-1}e^{2 \pi i \frac {\vec x^T \sigma \vec \xi}{\det L_2}}} {\det L_2M} I_2
\left(1-s, M^{-1}\det L_2, \vec x, \vec \xi\right)\\=
\frac {\pi^{2s-1}}{\det M \det^{2s}L_2}I_2
\left(1-s, M^{-1}, \vec x, \vec \xi\right)e^{2 \pi i \frac {\vec x^T \sigma \vec \xi}{\det L_2}}.
\end{multline}

Poisson formula implies $\hat I_2=I_2$. Hence (\ref{2d_eq}) is satisfied. $\blacksquare$

To find the Green function on $\mathbb {T}^2\times \mathbb{R}$ we add a non-periodic coordinate $y$,

 \begin{equation}
I_{cont}\left(s,\vec \xi, \vec x, y\right)=
c_0\sum_{\vec n \in \mathbb Z^2}\int dp\int \frac{dt}t t^s e^{-t\left({\vec x}^T+{\vec n}^TL_2\right)
\left({\vec x}+L_2^T{\vec n}\right)}e^{2\pi i {\vec n}^T L_2^\vee {\vec \xi}}e^{-tp^2}e^{2\pi i py}.
 \end{equation}

and use the (\ref{d2}) formula to get the integral equation for $I_{cont}$:

\begin{equation}
\label{article_eq}
I_{cont}\left(s,\vec \xi,\vec x, y\right)=\frac {\pi^{2s-\frac 32}}{\det^{2s-1}L_2}e^{2\pi i {\vec x}^T
\sigma \vec \xi}\int dp I_{cont} \left(\frac 32-s, \vec x, \vec \xi, \frac {p}{\det L_2}\right).
\end{equation}

This case is considered in details in the work \cite{article}.

{\bf Lemma 2.} \emph{The Green function for the Laplace equation on $\mathbb T^3$ satisfies the following functional equation:}

\begin{equation}
\label{fed3}
R_3\left(s,M,\vec x, \vec \xi\right)=
\frac{\Gamma\left(\frac 32 -s\right)\pi^{2s-\frac 32}}{\Gamma(s)\det L_3M}e^{-2\pi i \vec x^TL_3^{-1}
\left(L_3^T\right)^{-1}\vec \xi}R_3\left(\frac 32-s,\left(L_3^TL_3\right)^{-1}M,\vec\xi,\vec x\right).
\end{equation}

\underline{Proof.}

\begin{equation}\label{d3}\begin{array}{l}
\hat I_3\left(s,M,\vec \xi, \vec x\right)=c_0\sum_{\vec \nu \in \mathbb Z^3}\int d^3 n\int \frac{dt}t t^s e^{-t\left({\vec
x}^T+{\vec n}^TL_3\right)M^TM\left({\vec x}+L_3^T{\vec n}\right)}
e^{2\pi i {\vec n}^T L_3^{\vee T} {\vec \xi}}e^{-2\pi i {\vec \nu}^T \vec n}=\\
=c_0\sum_{\vec \nu \in \mathbb Z^3}\int \frac{dt} t t^s \left(\frac \pi t\right)^\frac 32 e^{-\frac{\pi^2}t\left({\vec
\xi}^TL_3^{\vee}\left(L_3^T\right)^{-1}-{\vec \nu}^T\left(L_3^T\right)^{-1}\right)
\left(M^{-1}\right)^TM^{-1}\left(L_3^{-1}L_3^{\vee T}\vec \xi-L_3^{-1}{\vec \nu}\right)}\\ e^{2\pi i {\vec
\nu}^T\left(L_3^T\right)^{-1}\vec x}e^{-2\pi i {\vec x}^T L_3^{-1}L_3^{\vee T}\vec \xi} \frac 1{\det {L_3M}}=\frac 1{\det
{L_3M}}c_0e^{-2\pi i {\vec x}^T L_3^{-1}L_3^{\vee T}\vec \xi}\\ \sum_{\vec \nu \in \mathbb Z^3} \int \frac{dt} t t^s
\left(\frac \pi t\right)^\frac 32 e^{-\frac{\pi^2}t\left({\vec \xi}^T-{\vec \nu}^TL_3\right)
\left(L_3^{-1}\left(L_3^T\right)^{-1}\right)M^TM\left(L_3^{-1}\left(L_3^T\right)^{-1}\right)\left(\vec \xi-L_3^T \vec
\nu\right)}
e^{2\pi i {\vec \nu}^TL_3^{\vee T}\vec x}=\\
=\frac{\pi^{2s-\frac 32}e^{-2\pi i {\vec x}^T L_3^{-1}\left(L_3^T\right)^{-1}\vec \xi}}{\det L_3M}I_3\left(\frac 32 -s,
L_3^{-1}\left(L_3^T\right)^{-1}M,\vec x, \vec \xi\right).
\end{array}\end{equation}

Once more applying the Poisson formula we obtain (\ref{fed3}). $\blacksquare$

We follow how the passage to the limit $c\to\infty$ changes the functional equation (\ref{fed3}) into the integral one (\ref{article_eq}). For convenience we take $c_x=c_y=0$, $M=\delta_{ij}$ and transform $L^\vee\to w_2L^\vee$ (see. (\ref{gen_w})), using the freedom of the definition of the $\vec n$ vector. Then $L_3^{-1}L_3^\vee=\begin{pmatrix} 0&-\frac 1{ab}&0\\ \frac 1{ab}&0&0\\ 0&0&\frac 1{c^2} \end{pmatrix}$ and (\ref{d3}) has the following form:

 \begin{multline}
\label{F_cont}
\hat I_{cont}\left(s,\vec \xi, \vec x\right)=\lim_{c\to\infty}
\left\{c_0\frac {\pi^{2s-\frac32}}{\det L_2 c}e^{2\pi i\frac{\vec \xi^T\sigma\vec x}{\det L_2}}
e^{-2\pi i \frac {z\zeta}{c^2}}\right.\sum_{\vec n}\sum_k e^{2\pi i \frac {zk}c}\\\left.\int
\frac {dt}t t^{\frac 32-s}e^{-\frac t{\det^2L_2}\left(\vec \xi^T+\vec n^TL_2\right)\left(\vec \xi+L_2\vec n\right)}
e^{2\pi i \vec n^TL_2^{\vee T}\vec \xi}e^{-\frac t{c^4}(\zeta+kc)^2}\right\}=
\lim_{c\to\infty}\left\{c_0\frac {\pi^{2s-\frac32}e^{2\pi i
\frac{\vec \xi^T\sigma\vec x}{\det L_2}}}{\det^{2s-4} L_2 }\right.\\\left.\sum_k
\frac {\Delta k}c\sum_{\vec n}\int\frac {dt}t t^{\frac 32-s}e^{-t\left(\vec \xi^T+\vec n^TL_2\right)
\left(\vec \xi+L_2\vec n\right)}e^{2\pi i \vec n^TL_2^{\vee T}\vec \xi}
e^{-\frac t{\det^2 L_2}(\frac \zeta{c^2}+\frac kc)^2}e^{2\pi i
\frac {zk}c}e^{-2\pi i \frac {z\zeta}{c^2}}\right\}=\\c_0\frac {\pi^{2s-\frac32}}{\det^{2s-4} L_2 }
e^{2\pi i\frac{\vec \xi^T\sigma\vec x}{\det L_2}}\sum_{\vec n}\int dp \int\frac {dt}t t^{\frac 32-s}
e^{-t\left(\vec \xi^T+\vec n^TL_2\right)\left(\vec \xi+L_2\vec n\right)}
e^{2\pi i \vec n^TL_2^{\vee T}\vec \xi}e^{-\frac {tp^2} {\det^2 L_2}}e^{2\pi i zp}\\=
c_0\frac {\pi^{2s-1}e^{2\pi i\frac{\vec \xi^T\sigma\vec x}{\det L_2}}}{\det^{2s-5} L_2 }
\sum_{\vec n}\int\frac {dt}t t^{\frac 32-s}e^{-t\left(\vec \xi^T+\vec n^TL_2\right)\left(\vec \xi+L_2\vec n\right)}
e^{2\pi i \vec n^TL_2^{\vee T}\vec \xi}e^{-\frac {\pi^2}t z^2 \det^2L_2}.
 \end{multline}

Here the three-dimensional parameter $\vec \xi$ is written as the two-dimensional vector $\xi$ and the coordinate along the non-periodical axis $\zeta$. We also pass to the limit $I_3\to I_{cont}$:

 \begin{multline}
\label{cont_def}
I_{cont}\left(s,\vec \xi, \vec x\right)=c_0\lim_{c\to\infty}\sum_{\vec n}\sum_k \int
\frac{dt}t t^s e^{-t\left({\vec x}^T+{\vec n}^TL_2\right)\left({\vec x}+L_2^T{\vec n}\right)}
e^{2\pi i {\vec n}^T L_2^{\vee T} {\vec \xi}}e^{-t(z+kc)^2}e^{2\pi i\frac {\zeta k}c}\\=
c_0\lim_{c\to\infty}\sum_{\vec n}\sum_{kc} \frac {\Delta kc}c \int \frac{dt}t t^s
e^{-t\left({\vec x}^T+{\vec n}^TL_2\right)\left({\vec x}+L_2^T{\vec n}\right)}
e^{2\pi i {\vec n}^T L_2^{\vee T} {\vec \xi}}e^{-t(z+kc)^2}e^{2\pi i\frac {\zeta}{c^2} kc}=\\
c_0\sum_{\vec n}\int dp\int \frac{dt}t t^s e^{-t\left({\vec x}^T+{\vec n}^TL_2\right)
\left({\vec x}+L_2^T{\vec n}\right)}e^{2\pi i {\vec n}^T L_2^{\vee T} {\vec \xi}}
e^{-tp^2}e^{2\pi i py}=\\c_0\sum_{\vec n}\int \frac{dt}t t^s e^{-t\left({\vec x}^T+{\vec n}^TL_2\right)
\left({\vec x}+L_2^T{\vec n}\right)}e^{2\pi i {\vec n}^T L_2^{\vee T} {\vec \xi}}e^{-\frac {\pi^2}ty^2}.
 \end{multline}

Here we cannot use the Poisson formula $I=\hat I$ literally because there is no lattice in the $z$ direction. This means that we actually have performed an extra Fourier transformation along $p$ while computing (\ref{d3}) in the limit $c\to\infty$. Taking this into consideration and putting (\ref{F_cont}) and (\ref{cont_def}) together we obtain the integral equation, 

 \begin{equation}
\int dp I_{cont}(s, \vec \xi, \vec x, p) e^{-2\pi i py}=
\frac {e^{2\pi i\frac{\vec \xi^T\sigma\vec x}{\det L_2}}}{\sqrt{\pi}\det L_2}
\frac {\pi^{2s-1}}{\det^{2s-5} L_2 } I_{cont}\left(\frac 32-s, \vec x, \vec \xi, y\det L_2\right).
 \end{equation}

The factor $\frac 1{\sqrt{\pi}\det L_2}$ arises from the inverse Fourier transformation. We substitute $s\to \frac 32-s$. Then,

 \begin{equation}
\label{fed2c}
I_{cont}\left(s,\vec \xi,\vec x, y\right)=
\frac {\pi^{2s-\frac 32}}{\det^{2s-1}L_2}e^{2\pi i {\vec x}^T\sigma \vec \xi}\int dp I_{cont}
\left(\frac 32-s, \vec x, \vec \xi, \frac {p}{\det L_2}\right).
 \end{equation}

This expression coincides with (\ref{article_eq}). Thus we have obtained a well-defined expression for the Green function of the Laplace equation on the three-dimensional torus,

 \begin{equation}
\label{expl}
\varphi\left(\vec x, \vec \xi \right)=e^{-2\pi i \vec x^T L_3^{-1}L_3^{\vee T}\vec \xi}
\sum_{\vec n}\frac {e^{2\pi i \vec n^TL_3^{\vee T}\vec x}}{\left(\vec \xi^T+\vec n^TL_3\right)\left((L_3^\vee)^TL_3^\vee\right)^2
\left(\vec \xi+L_3^T\vec n\right)}.
 \end{equation}
\section{Modular properties}
\label{mod}
\subsection{Complex case}

Consider the transformation of the Green function on the three-dimensional torus under the action of the element from the $SL(3,\mathbb Z)\backslash SL(3,\mathbb R)/SO(3,\mathbb R)$ coset. The $SL(3,\mathbb Z)$ group acts as modular transformations of the $\Gamma_3$ lattice. The $SO(3, \mathbb R)$ group allows to choose an orthonormal basis for the lattice vectors, i.e. in our case to transform the matrix to lower triangular form. For example consider the action of an element of the double coset $SL(2,\mathbb Z)\backslash SL(2,\mathbb R)/SO(2,\mathbb R)$ on the Green function on $\mathbb T^2 \times \mathbb R$. We write the non-regularized solution in the complex form. Let $a=1$, $\tau=b_x+ib$, and the lattice vectors $\gamma=n+m\tau$. The area of the elementary cell is $S=\Im \tau$. Then the formula for the character (\ref{chi}) can be written as:

\begin{equation}
\chi(\gamma,\xi)=\exp\left\{\frac \pi S \left(\gamma \bar \xi-\bar \gamma \xi \right)\right\},
\end{equation}

and the expression (\ref{R_def}) has the following analogue:

\begin{equation}
\label{R_comp}
R(s,\xi,w,y)=c_0\sum_\gamma \frac{\chi(\gamma,\xi)}{\left(\left(yS\right)^2+|w+\gamma|^2\right)^s}.
 \end{equation}
 
To study the modular properties of the solution (\ref{R_comp}) it is sufficient to see how it changes under the transformations $\tau \to \tau +1$ and $\tau \to -1/\tau$. In the first case the summation over $n$ in the solution (\ref{R_comp}) is shifted: $n\to n+m$. Hence the solution remains invariant. In the second case the transformation acts a bit more complicated,

\begin{equation}
\begin{array}{l}
\tau\to-\frac 1\tau=-\frac {\bar \tau}{\tau \bar\tau},\ \ S\to \frac S {\tau\bar\tau},\\
\exp\left\{\frac \pi S \left(\gamma \bar \xi-\bar \gamma \xi \right)\right\}\to
\exp\left\{\frac \pi S \left(\left(n\tau-m \right)\bar \tau \bar \xi - \left(n\bar \tau-m \right)\tau\xi\right)\right\},\\
(pS)^2+|w+\gamma|^2\to\frac 1 {\tau\bar\tau} \left(\left(\frac p {\sqrt {\tau\bar\tau}}S\right)^2+ |w\tau+n\tau-m|^2\right).
\end{array}
 \end{equation}
 
Summing over $n\tau -m$ instead of $n+m\tau$ we find

\begin{equation}
\label{mod_comp}
R(\frac 12, \xi, w, p)\to \sqrt{\tau \bar \tau}R(\frac 12, \tau \xi, \tau w, \frac p {\sqrt {\tau\bar\tau}}).
 \end{equation}
 
To understand how the connections are transformed we substitute $\sqrt{\tau\bar\tau}\phi\left(\tau z, \tau \bar z, \frac y{\sqrt{\tau\bar\tau}}\right)$ instead of $\phi\left(z,\bar z,y\right)$ in the Bogomolny equation (\ref{sys_t2}). For the system to remain invariant the following substitutions are needed:
 
\begin{equation}
\label{Mod_system}
\left\{
\begin{array}{l}
A_z\left(z,\bar z,y\right)\to \tau A_z\left(\tau z,\bar \tau \bar z, \frac y{\sqrt{\tau\bar\tau}}\right),\\
A_{\bar z}\left(z,\bar z,y\right)\to \bar \tau A_{\bar z}\left(\tau z,\bar \tau \bar z, \frac y{\sqrt{\tau\bar\tau}}\right),\\
A_y\left(z,\bar z,y\right)\to \frac 1{\sqrt{\tau\bar\tau}} A_y\left(\tau z,\bar \tau \bar z, \frac y{\sqrt{\tau\bar\tau}}\right).
\end{array}
\right.
 \end{equation}
 
Hence the connections $A_zdz$, $A_{\bar z}d\bar z$, $A_ydy$ are invariant under modular transformations. The modular-transformed solution could be expected to differ from the initial one by the gauge, but it is not so: the invariance of the solution means implies the absence of the holomorphic function satisfying the quasi-periodicity condition (\ref{cond}).

\subsection{Three-dimensional case}

\begin{figure}
\begin{tabular}{ccccccc}
\parbox[c]{41 pt}{\includegraphics[angle=-90, width=41 pt]{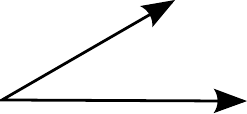}}
&
\parbox[c]{13 pt}{\includegraphics[angle=-90, width=13 pt]{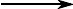}}
&
\parbox[c]{70 pt}{\includegraphics[angle=-90, width=70 pt]{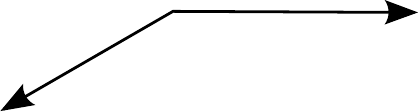}}
&
\parbox[c]{13 pt}{\includegraphics[angle=-90, width=13 pt]{05}}
&
\parbox[c]{68 pt}{\includegraphics[angle=-90, width=68 pt] {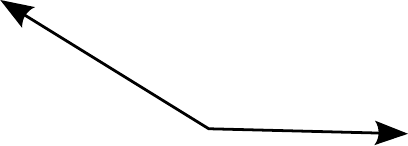}}
&
\parbox[c]{13 pt}{\includegraphics[angle=-90, width=13 pt]{05}}
&
\parbox[c]{79 pt}{\includegraphics[angle=-90, width= 79 pt]{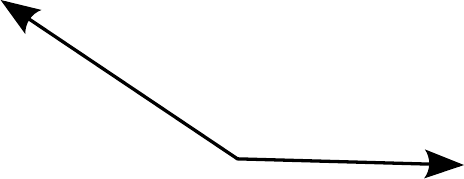}}\\[20 pt]
$(1,\tau)$&$\rightarrow$ &$(-\tau,1)$&$\rightarrow$ &$(\tau\cdot\frac{\bar\tau}{\sqrt{\tau\bar\tau}},
-\frac{\bar\tau}{\sqrt{\tau\bar\tau}})$&$\rightarrow$ &$(1,-\frac 1 \tau)$\\[20 pt]
$\begin{pmatrix}1&0\\b_x&b\end{pmatrix}$&$\rightarrow$ &$\begin{pmatrix}-b_x&-b\\1&0\end{pmatrix}
$&$\rightarrow$ &$\begin{pmatrix}\sqrt{b_x^2+b^2}&0\\-\frac {b_x}{\sqrt{b_x^2+b^2}}&\frac b{\sqrt{b_x^2+b^2}}\end{pmatrix}
$&$\rightarrow$ &$\begin{pmatrix}1&0\\-\frac{b_x}{b_x^2+b^2}&\frac b{b_x^2+b^2}\end{pmatrix}$\\
\end{tabular}
\caption{The modular transformation of a pair of vectors in complex and matrix forms.}\label{fig}
 \end{figure}

To understand how the solution is transformed under the action of the modular group we consider the transformation of the lattice and character matrices. The left action of the $SL(3,\mathbb Z)$ generators on the lattice matrix $L$ permutes the rows of the matrix and changes their signs; therefore $L$ ceases to be lower triangular. Right action of the $SO(3,\mathbb R)$ allows to make $L$ lower triangular and multiplication by a constant allows to restore the $a$ matrix element in the upper left angle. This procedure for the two-dimensional case is depicted in fig.\,\ref{fig}.

Hence $L$ transforms as following:

 \begin{equation}
\label{mod_L}
L\to\frac 1{\alpha_i}w_iLO_i,\ \ O_i\in SO(3,\mathbb R), \ \alpha_i \in \mathbb R.
 \end{equation}
 
Using (\ref{mod_L}) and (\ref{Lvee_def}) we derive the transformation law for the dual matrix:

 \begin{equation}
L^\vee\to\alpha_i w_i L^\vee O_i.
 \end{equation}
 
The modular transformation of the solution (\ref{R_def}) in general can be expressed through $O_i, \alpha_i$:

 \begin{multline}
c_0\sum_{n\in\mathbb Z^3}\frac{e^{2\pi i \vec n^TL^\vee\vec \xi}}{\left(\left(\vec x^T+\vec n^TL\right)MM^T
\left(\vec x+L^T\vec n\right)\right)^s}\to \\c_0\sum_{n\in\mathbb Z^3}\frac{e^{2\pi i
\vec n^T w_i L^\vee \alpha_i O_i\vec \xi}}{\left(\left(\vec x^T+\vec n^Tw_iL\frac 1 \alpha_i O_i\right)MM^T
\left(\vec x+\frac 1 \alpha_i O_i^T L^T w_i^T\vec n\right)\right)^s}=\\
c_0\sum_{\nu\in \mathbb Z^3}\frac{\alpha_i^{2s}e^{2\pi i \vec \nu^T L^\vee
\left(\alpha_i O_i\vec \xi\right)}}{\left(\left(\left(\alpha_i\vec x^TO_i^{-1}\right)+\vec \nu^TL\right)O_iMM^TO_i^T
\left(\left(\alpha_i O_i\vec x\right)+ L^T \vec \nu\right)\right)^s},\ \ \ \nu=w_i^T\vec n,
 \end{multline}
 
Hence,

 \begin{equation}
  R(s,M,\vec x, \vec \xi)^{w_i}=\alpha_i^{2s}R(s,O_iM,\alpha_iO_i\vec x, \alpha_iO_i\vec \xi).
 \end{equation}
 
We compute $\alpha_i$ and $O_i$ for all $w_i$. $w_1$ and $w_3$ shift the summation (similarly to $\tau \to \tau
+1$ in two-dimensional case): the first one corresponds to the shift $n\to n+m$, the second one corresponds to the change of signs of $m$ and $k$. So $w_1$ and $w_3$ don't change the solution. $w_2$ acts more complicated:

 \begin{multline}
\begin{pmatrix} a&0&0\\b_x&b&0\\c_x&c_y&c\end{pmatrix}\to \frac 1 \alpha_2
\begin{pmatrix}0&1&0\\-1&0&0\\0&0&1\end{pmatrix}\begin{pmatrix} a&0&0\\b_x&b&0\\c_x&c_y&c\end{pmatrix}
\begin{pmatrix} \frac {b_x}{\sqrt{b^2+b_x^2}}&-\frac{b}{\sqrt{b^2+b_x^2}}&0\\\frac{b}{\sqrt{b^2+bx^2}}&
\frac{b_x}{\sqrt{b^2+b_x^2}}&0\\0&0&1\end{pmatrix}=\\\frac 1 {\alpha_2}\frac 1{\sqrt{b^2+b_x^2}}
\begin{pmatrix}{b^2+b_x^2}&0&0\\-ab_x&ab&0\\bc_y+b_xc_x&b_xc_y-bc_x&c\sqrt{b^2+b_x^2}\end{pmatrix}.
 \end{multline}

So $\alpha_2=\frac{\sqrt{b^2+b_x^2}}a=\frac {|\gamma_2|}{|\gamma_1|}$ and $O_2=T_z^\alpha$ where $\alpha=\angle(\vec \gamma_1, \vec \gamma_2)$. Here $T_z^\alpha$ is a rotation matrix around $z$ axis by $\alpha$ angle. In the case of $w_4$ matrix the expression for the rotation matrix is quite bulky. We write $O_4$ as a composition of three matrices:

 \begin{equation}
O_4=T^\alpha_z T^\beta_y T^\gamma_x
\end{equation}
\begin{equation}
\alpha=\angle(\vec \gamma_1, \vec \gamma_3), \ \ \beta=
\frac \pi 2 - \angle(\vec \gamma_3, \vec z),\ \ \tan\gamma=\sin \beta \cot \angle(\vec \gamma_2,\vec \gamma_3),\,\,\,\alpha_4=
\frac{|\gamma_3|}{|\gamma_1|}.
 \end{equation}
 
Modular transformations multiply the solution by a constant which in the corresponding basis of $SL(3,\mathbb Z)$ is a ratio of lengths of the lattice vectors. Substituting the transformed solution into (\ref{Bogomolny}) we derive the invariance of the connection under modular transformations.

{\bf Acknowledgments.} The author is grateful to M.\,A.\,Olshanetsky, A.\,M.\,Levin and A.\,Zotov for suggesting the problem and fruitful discussions. The work was supported with the grant FASI RF 14.740.11.0347.

\section{\appendixname}

The character of the torus is the multiplicative periodic function on the plane. Its periods are the generating vectors of the lattice, $\vec \gamma_i \in \Gamma_d,$:

 \begin{equation}
\label{char}
\begin{array}{l}
\chi\left(\vec x+\vec y,\vec n\right)=\chi\left(\vec x, \vec n\right)\chi\left(\vec y, \vec n\right),\ \ \ \vec n\in \mathbb{Z}^d,\\
\chi\left(\vec \gamma_i, \vec n\right)=1.
\end{array}
 \end{equation}
 
In general this function is expressed through  the vectors of the dual lattice:

 \begin{equation}
\label{chi}
\chi\left(\vec x, \vec n\right)=\exp\left\{2 \pi i \left(\vec x \vec \alpha\right)\right\},\,\,\,
\left(\vec \alpha_i \vec \gamma_j\right)=\delta_{ij}.
 \end{equation}

Here  $\alpha_i$ are the generators of the dual lattice. Introduce the matrices of the lattice and the dual lattice,

 \begin{equation}
\begin{array}{l}
\gamma=L\cdot \vec n,\,\,\gamma\in \Gamma_d,\,\,\,\vec n \in \mathbb{N}^d,\\
\alpha=L^{\vee}\cdot \vec n,\,\,\gamma\in \Gamma^{\vee}_d,\,\,\,\vec n \in \mathbb{N}^d,
\end{array}
 \end{equation}
 
where $d$ is the dimension of the manifold. The matrices of the dual lattices can be found from the expressions:

 \begin{equation}
\label{Lvee_def}
\begin{array}{l}
L_2^{\vee}=L_2^{-1}\sigma,\,\,\,\sigma=\begin{pmatrix}0 & 1 \\-1 & 0\end{pmatrix},\\
L_3^{\vee}=L_3^{-1}.
\end{array}
 \end{equation}

For calculations we need the explicit form of these matrices. Choose the lattice matrices in the lower triangular form,

 \begin{equation}
\label{l2}
L_2=\begin{pmatrix}
a & 0 \\
b_x & b
\end{pmatrix},\,\,\,
L_3=\begin{pmatrix}
a & 0 & 0 \\
b_x & b & 0 \\
c_x & c_y & c
\end{pmatrix},
 \end{equation}

Write the matrices of the dual lattices,

 \begin{equation}
\label{vee}
L_2^\vee=\frac 1{ab}\begin{pmatrix}
0 & b \\
-a & -b_x
\end{pmatrix},\,\,\,
L_3^\vee=\frac 1{abc} \begin{pmatrix}
bc & 0 & 0 \\
-b_xc & ac & 0 \\
b_xc_y-bc_x & -ac_y & ab
\end{pmatrix}
 \end{equation}

The generators of  $SL(3, \mathbb Z)$ can be chosen as follows \cite{discrete}:

 \begin{equation}
\label{gen}
U=\begin{pmatrix}1&1&0\\0&1&0\\0&0&1\end{pmatrix},\ \
P=\begin{pmatrix}0&-1&0\\-1&0&0\\0&0&-1\end{pmatrix}, \ \
Q=\begin{pmatrix}0&1&0\\0&0&1\\1&0&0\end{pmatrix}, \ \ O=\begin{pmatrix}1&0&0\\0&-1&0\\0&0&-1\end{pmatrix}
 \end{equation}

It is more convenient to use another choice of the generators,

\begin{equation}
\label{gen_w} w_1=U,\ \ w_2=PO=\begin{pmatrix}0&1&0\\-1&0&0\\0&0&1\end{pmatrix},\ \ w_3=O,\ \
w_4=POQ=\begin{pmatrix}0&0&1\\0&-1&0\\1&0&0\end{pmatrix}.
\end{equation}

\end{document}